
\catcode`@=11 
\def\binrel@#1{\setbox\z@\hbox{\thinmuskip0mu
 \medmuskip-1mu\thickmuskip\@ne mu$#1\m@th$}%
 \setbox\@ne\hbox{\thinmuskip0mu\medmuskip-1mu\thickmuskip
 \@ne mu${}#1{}\m@th$}%
 \setbox\tw@\hbox{\hskip\wd\@ne\hskip-\wd\z@}}
\def\binrel@@#1{\ifdim\wd2<\z@\mathbin{#1}\else\ifdim\wd\tw@>\z@
 \mathrel{#1}\else{#1}\fi\fi}
\def\pmb{\let\next\relax\ifmmode\def\next{\mathpalette\pmb@}\else
 \let\next\pmb@@\fi\next}
\def\pmb@@#1{\leavevmode\setbox\z@\hbox{#1}\kern-.025em\copy\z@\kern-\wd\z@
 \kern-.05em\copy\z@\kern-\wd\z@\kern-.025em\raise.0433em\box\z@}
\newdimen\pmbraise@
\def\pmb@#1#2{\setbox\thr@@\hbox{$\m@th#1{#2}$}%
 \setbox4\hbox{$\m@th#1\mkern.7794mu$}\pmbraise@\wd4
 \divide\pmbraise@18
 \binrel@{#2}\binrel@@{\mkern-.45mu\copy\thr@@\kern-\wd\thr@@
 \mkern-.9mu\copy\thr@@\kern-\wd\thr@@\mkern-.45mu\raise\pmbraise@\box\thr@@}}
\catcode`@=12 

\def\lsim{\mathrel{\rlap{\lower4pt\hbox{\hskip1pt$\sim$}}
    \raise1pt\hbox{$<$}}}         
\def\gsim{\mathrel{\rlap{\lower4pt\hbox{\hskip1pt$\sim$}}
    \raise1pt\hbox{$>$}}}         

\def\overleftrightarrow#1{\vbox{\ialign{##\crcr
    $\leftrightarrow$\crcr
    \noalign{\kern 1pt\nointerlineskip}
    $\hfil\displaystyle{#1}\hfil$\crcr}}}
    \long\def\caption#1#2{{\setbox1=\hbox{#1\quad}\hbox{\copy1%
    \vtop{\advance\hsize by -\wd1 \noindent #2}}}}

\def\frac#1/#2 {{\textstyle {#1\over #2}}}

\def\dag{\dagger}

\def\s0{\sigma_0}

\def\underrightarrow#1{\vtop{\ialign{##\crcr
    $\hfil\displaystyle{#1}\hfil$\crcr\noalign{\kern0pt\nointerlineskip}
     \rightarrowfill\crcr\noalign{\kern0pt}}}}

\normalbaselineskip=18pt
\baselineskip=18pt

\overfullrule=0pt

\normalbaselineskip=12pt
\baselineskip=12pt
\magnification=1200
\hsize 5.0 in
\vsize 7.5 in
\line{\hfil DOE/ER/40427-31-N93}
\vskip 36pt
\centerline{\bf A QUARK MODEL OF $\pmb{\bar \Lambda \Lambda}$ PRODUCTION OF
$\pmb{\bar p p}$
INTERACTIONS}
\vskip 18pt
\centerline{M.A. Alberg$^{(1), (2)}$, E.M. Henley$^{(2)}$, L. Wilets$^{(2)}$,
and P.D.
Kunz$^{(3)}$}
\medskip
\centerline{\it $^{(1)}$ Department of Physics, Seattle University, Seattle,
Washington 98122
USA}
\centerline{\it $^{(2)}$ Department of Physics, FM-15, University of
Washington,
 Seattle, WA  98195, USA}
\centerline{\it $^{(3)}$ Nuclear Physics Laboratory, University of Colorado,
Boulder, Colorado
80309 USA}
\vskip2.0truein
\centerline{To be published in ``Journal of Atomic Nuclei"}
\centerline{Paper given at ``2nd Biennial
Workshop on Nuclear - Antinuclear Physics"}
\centerline{ITEP, Moscow, Russia, 13th - 18th Sept. 1993.}
\vskip1.0truein  \centerline{PREPARED FOR THE U.S. DEPARTMENT OF ENERGY}
\vskip 6pt
\normalbaselineskip=10pt
\baselineskip=10pt

\noindent This report was prepared as an account of work sponsored by the
United States Government.  Neither the United States nor the
United States Department of Energy, nor any of their employees,
nor any of their contractors, subcontractors, or their employees,
makes any warranty, express or implied, or assumes any legal
liability or responsibility for the product or process disclosed,
or represents that its use would not infringe privately-owned
rights.
\vskip 6pt
\noindent By acceptance of this article, the publisher and/or recipient
acknowledges the U.S. Government's right to retain a nonexclusive,
royalty-free license in and to any copyright covering this paper.

\vfill\eject

\magnification=1200
\normalbaselineskip=24truept
\baselineskip=24truept
\overfullrule=0pt
\hsize 16.0true cm
\vsize 22.0true cm

\def\lsim{\mathrel{\rlap{\lower4pt\hbox{\hskip1pt$\sim$}}
    \raise1pt\hbox{$<$}}}         
\def\gsim{\mathrel{\rlap{\lower4pt\hbox{\hskip1pt$\sim$}}
    \raise1pt\hbox{$>$}}}         

\def\overleftrightarrow#1{\vbox{\ialign{##\crcr
    $\leftrightarrow$\crcr
    \noalign{\kern 1pt\nointerlineskip}
    $\hfil\displaystyle{#1}\hfil$\crcr}}}
\long\def\caption#1#2{{\setbox1=\hbox{#1\quad}\hbox{\copy1%
\vtop{\advance\hsize by -\wd1 }}}}
\centerline{\bf  A QUARK MODEL OF $\pmb{\bar \Lambda \Lambda}$}
\centerline{\bf PRODUCTION IN $\pmb{\bar p p}$  INTERACTIONS}
\medskip
\centerline{M.A. Alberg$^{(1),(2)}$, E.M. Henley$^{(2)}$, L. Wilets$^{(2)}$ and
P.D.
Kunz$^{(3)}$}
\smallskip
\centerline{\it $^{(1)}$Department of Physics, Seattle University, Seattle,
Washington
98122 USA}
\centerline{\it $^{(2)}$Department of Physics, University of Washington,
Seattle,
Washington 98195 USA}
\smallskip
\centerline{\it $^{(3)}$Nuclear Physics Laboratory, University of Colorado,
Boulder,
 Colorado  80309 USA}
\vskip 18 pt
\centerline{\bf ABSTRACT}
\rm A quark model which includes both scalar and vector
contributions to the 	reaction mechanism (SV quark model) is used in a DWBA
calculation of $\bar \Lambda \Lambda$ production in $\bar p p$ interactions.
Total and
differential cross-sections, polarizations, depolarizations, and
spin-correlation coefficients
are computed for laboratory momenta from threshold to 1695 MeV/c.  The free
parameters of the
calculation are the scalar and vector strengths, a quark cluster size
parameter,
and the
parameters of the unknown $\bar \Lambda \Lambda$ potentials.  Good agreement
with experiment is
found for constructive interference of the scalar and vector terms, and for
$\bar \Lambda \Lambda$ potentials which differ from those suggested by several
authors on the
basis of SU(3) arguments.  The fit to the data is better than that obtained by
other quark
models, which use only scalar $or$ vector annihilation terms.  The agreement
with experiment
is also better than that found in meson-exchange models.  The recent suggestion
[1] that
measurement of the depolarization parameter $D_{nn}$ can be used to
discriminate
between
meson-exchange and quark models is examined in detail.  We conclude that a
measurement of
$D_{nn}$ will provide a test of which of these models, as presently
constructed,
is the more
appropriate description of strangeness production in the $\bar p p \rightarrow
\bar \Lambda
\Lambda$ reaction. \medskip
\noindent
Speaker:  M.A. Alberg, Department of Physics, Seattle University, Seattle WA
98122, USA..
E-mail:  alberg@uwaphast.bitnet.  FAX:  (206) 685-0635.
\vfill \eject
\centerline{\bf 1. Introduction}
\medskip
We describe the results of a DWBA calculation of
the total and differential cross sections, polarizations, and spin-correlation
coefficients that have been measured by the PS185 collaboration [2] for the
reaction $\bar p p
\rightarrow \bar \Lambda \Lambda$ from threshold to
1695 MeV/c. We also present predictions for the proposed measurement [3] of the
depolarization parameter $D_{nn}$.  The $\bar p p
\rightarrow \bar \Lambda \Lambda$ reaction can be described in terms of either
quark or meson-exchange models, and may provide a test of which picture is more
appropriate at the momenta and distances which correspond to the
experimental measurements. Because the reaction is very sensitive
to initial and final state interactions it also can provide information
about the $\bar \Lambda \Lambda$ interaction, for which there are no direct
experimental measurements.

Several groups have used meson-exchange models [4-9] to
obtain reasonable fits to the data. The $K$, $K^*$ and $K^{**}$ exchanges in
these
models are of  short range, at  distances
for which one might expect quark degrees of freedom to be
important.  Quark models provide a microscopic picture of the
reaction which tests our understanding of non-perturbative QCD. In
the simplest quark models either a scalar ($``^3P_0"$) or vector ($``^3S_1"$)
interaction is assumed to describe $\bar q q$ annihilation and creation, and
several calculations [6,10-15] have obtained reasonable agreement
with experiment. In some cases these results have been used to
argue that either the scalar or vector interaction provides the
correct description of annihilation. We have proposed [16] that both
mechanisms should be included, since by analogy with the $N N$
interaction one would expect at least vector exchange (of one or more
gluons ) and a scalar representation of both confinement and
multigluon exchange.

In theoretical calculations to date, quark and meson-exchange models have been
about equally
successful in fitting experimental data, although our SV quark model is better
at
reproducing the steep rise seen in the differential cross-section at forward
angles.  Recently
the J\"ulich group has proposed [1] that the depolarization parameter $D_{nn}$
could be used
to discriminate between the quark and meson-exchange models.  The quark model
they used was the
vector mechanism proposed by Kohno and Weise [6].  We have carried out
calculations of $D_{nn}$
for our SV model, and find that even with the inclusion of the scalar term in
the reaction
mechanism, the quark model predictions differ strongly from those of
meson-exchange.

Sections 2 through 5 provide a summary of our quark model calculation,
including
comparison
with experimental results and with other quark and meson-exchange models.  A
complete
description of the calculation, together with comparisons to experimental data
of PS185
for the $\bar p p \rightarrow \bar \Lambda \Lambda$ reaction, has recently been
published [17]. In section 6 we compare our SV quark model predictions for
$D_{nn}$ to the
meson-exchange calculations of the J\"ulich group.

\bigskip
\centerline{\bf 2. Reaction Mechanism}
\medskip
Our reaction mechanism
includes both scalar and vector contributions to the annihilation and creation
of
antiquark-quark  pairs. The simplest graphs for these terms are shown in Fig.
1.
The  $``^3P_0"$
term represents scalar multigluon exchange and/or the  confining scalar force,
whereas the
$``^3S_1"$ term represents vector  exchange of one or more gluons. Both terms
also include $\bar
q q$  pairs in  intermediate states. In our model, the operator for scalar
exchange is
zero-range and of the form
 $$I_s = g_s {\pmb\sigma'_3} \cdot \left({\pmb\nabla_{3'} -
\pmb\nabla_{6'} \over 2m_s} \right) \pmb\sigma_3 \cdot \left( {\pmb\nabla_3 -
\pmb\nabla_6 \over 2m} \right) \,, \eqno(1)$$
and that for vector exchange is
$$I_v = g_v {\pmb\sigma'_3} \cdot {\pmb\sigma_3 }\eqno(2)$$
In (1), $m_s$ is the strange quark mass and $m$ is the up quark mass. For $\bar
\Lambda
\Lambda$ production the active quarks are a $\bar u u$   pair which is
annihilated and an
$\bar s s$  pair which is created. The  spectator quark pairs, $\bar u \bar d$
and $ud$, must
each be in an I=0 and S=0  state, so that the spin of the $\bar \Lambda
\Lambda$
pair is
carried by the strange  quarks. Both scalar and vector terms are spin triplet,
so our
quark model predicts a singlet fraction for the  $\bar \Lambda \Lambda$ pair
which is  identically zero, in good agreement with experiment.  A singlet
contribution can
arise, e.g. from a pseudoscalar $``^1S_0"$  term, but this  has been shown to
be
small [18].
\bigskip
\centerline{\bf 3. Initial and Final State
Interactions} \bigskip \noindent 3.1. THE $\bar p p$  INTERACTION
\medskip
In  most of our work we use the $\bar p p$ potential proposed by Kohno and
Weise [6]
$$V_{\bar N N} (r) = U_{\bar N N} (r) + i W_{\bar N N} (r)\eqno(3)$$
in which the real term $U_{\bar N N} (r)$ includes central, tensor, spin-orbit
and
spin-spin terms and the imaginary term $W_{\bar N N} (r)$ is a central
potential
which represents annihilation. The long-range part of $U_{\bar N N} (r)$ is
determined by the G-parity transform of Ueda's [19] one-boson
exchange potential. For r $<$ 1 fm each term
in the real part of the potential is extrapolated smoothly to the origin by
means of a
Woods-Saxon form. The imaginary potential $W_{\bar N N} (r)$ is given by:
$$W_{\bar N N} (r) = W_{\bar N N}^{(0)} \left\{ 1 + \exp [(r-r_0)/a]
\right\}^{-1} \eqno(4)$$
\noindent
with $r_0$ = 0.55 fm, a = 0.2 fm, and $W_{\bar N N}^{(0)} = -1.2$ GeV. These
parameters
give good fits to total, elastic, annihilation and charge-exchange $\bar p p$
data  for lab momenta up to 2.5 GeV/c.
\bigskip
\noindent 3.2. THE $\bar \Lambda \Lambda$ INTERACTION
\medskip
Our  $\bar \Lambda \Lambda$ potential is chosen to be of the form used by Kohno
and Weise [6], although we find it necessary to vary the parameters
of that potential to get good agreement with experiment. In the Kohno-Weise
$\bar \Lambda
\Lambda$ potential $$V_{\bar \Lambda \Lambda}(r) =
U_{\bar \Lambda \Lambda}(r) +iW_{\bar \Lambda \Lambda}(r)
\eqno(5)
$$
the real term  $U_{\bar \Lambda \Lambda} (r)$ represents isoscalar meson
exchange
and the imaginary term  $W_{\bar \Lambda \Lambda} (r)$ represents annihilation.
The
long-range part of $ U_{\bar \Lambda \Lambda} (r)$ is derived from the
isoscalar
exchanges of
the  Nijmegen YN potential [20], in which SU(3) relations were used to
determine couplings
for the pseudoscalar, vector, and scalar nonets. As in the $\bar N N$ case, the
short-range
part is determined by means of a smooth  extrapolation to the origin. The
imaginary term
W$_{\bar \Lambda \Lambda} (r)$ is taken to be a Woods-Saxon form with the same
radius and
diffuseness as the $\bar N N$ absorptive potential:
$$W_{\bar \Lambda \Lambda} (r) = W_{\bar \Lambda \Lambda}^{(0)} \left\{ 1 +
\exp
[(r-r_0)/a]
\right\}^{-1} \eqno(6)$$
\noindent
The strength $W_{\bar \Lambda \Lambda}^{(0)}$ = -700 MeV was chosen to fit the
$\bar \Lambda
\Lambda$ production  cross-section.
The Kohno-Weise potential described above is based in part on
SU(3) symmetry arguments together with the use of the G-parity
transformation, both of which may be questioned. In the absence of
direct experimental data on the $\bar \Lambda
\Lambda$ interaction, this potential is a
good starting point for our analysis. But as we describe in the next
section, good fits to the experimental data require changes in the
parameters of the potential, and thereby give us information about
the $\bar \Lambda
\Lambda$ interaction that has not been previously available.
\bigskip
\centerline{\bf 4. Comparison with Experiment}
\medskip
\noindent
A nine-parameter fit was made to the PS185 data reported at lab momenta of
1436,
1437, 1445,
1477, 1508, 1546, 1642 and 1695  MeV/c. The 356 data points included in this
set
of
measurements  include differential cross sections, polarizations and spin
correlation coefficients. No data points were excluded from the fits.

Minimization programs which included Monte Carlo, simplex, gradient, and
simulated annealing
techniques were used to search  for the best values of $g_v$  (the strength
of the vector term), $g_s$ (the strength of the scalar  term), $r_0$ (a range
parameter in
the quark Gaussian wave function),  and six parameters in the $\bar \Lambda
\Lambda$ potential. The searches were started with the Kohno-Weise values for
the
$\bar\Lambda\Lambda$  potential parameters. Three parameters for the real  part
of the potential
were varied: $V$ (the strength of the central  plus spin-spin term), $V_T$
(the
strength of the
tensor term), and $V_{LS}$ (the strength of the spin-orbit term). Three
parameters were varied
in the annihilation term: $W_{\bar \Lambda \Lambda}^{(0)},\,\, r_w$ and $a_w$
(the strength,
radius and  diffuseness of the potential). 	We considered three possible
cases for the reaction mechanism:  the vector  $``^3S_1"$ term alone, the
scalar $``^3P_0"$ term alone, and a superposition of both terms. The
results of our searches are shown in Table 1.  All of our searches found minima
for
very small values of $a_w$, the diffuseness of the annihilation potential. We
therefore fixed
$a_w$ at the value of 0.01 fm to avoid reflections from a sharp square-well
potential, which
left 8 free parameters to be varied in our searches.  Clearly the
superposition
of both terms
provides the best fit to the data, with a  $\chi^2$ per data point of 3.2.  An
example of the
quality of our fit to the data using our best global fit parameters is shown in
Fig. 2. A complete comparison with all experimental data is given in reference
17. The best
global fit parameters given in Table 1 indicate that the  $\bar \Lambda
\Lambda$
potential
differs from that expected on the basis of SU(3)  arguments. In the real part
of
the potential,
for which the long-range  behavior is determined by one-boson exchange, the
central term is
small. The tensor and
spin-orbit terms are much larger than the predictions of the
one-boson exchange model. This may reflect a greater spin-dependence in the
interaction,
which has also been noted by other investigators [7-9].  The annihilation term
in the  $\bar
\Lambda \Lambda$ potential  is deeper, longer in range, and much less diffuse
than the
Kohno-Weise annihilation potential.

	In Fig. 3 we show a comparison of the best vector alone, scalar
alone, and combined reaction mechanism calculations with the
experimental data at 1642 MeV/c. Here it is seen that both terms
are needed in the reaction mechanism to get a good fit to both the
differential cross section and polarization data. These results are
representative of all the laboratory momenta studied. Vector or
scalar terms alone can fit the differential cross sections reasonably well, but
only a superposition fits the spin observables as well.
\bigskip
\centerline{\bf 5. Comparison to other Theoretical Calculations}
\bigskip
\noindent
5.1 QUARK MODELS
\medskip
Quark model calculations of differential cross sections and polarizations have
been made by
Kohno and Weise [6] and Furui and Faessler [14].  Kohno and Weise used the
vector
``$^3S_1$" model with the initial and final state interactions we have
described
above.  Furui
and Faessler considered separately the vector ``$^3S_1$" and the scalar
``$^3P_0$" models, with
initial and final state interactions taken from the meson-exchange calculations
of Tabakin and
Eisenstein [4].  They concluded that their ``$^3P_0$" model was in better
agreement with the
data.  The results of both calculations are compared to ours and to the data in
Fig.~4.  Our
calculations are in better agreement with the differential cross sections.  The
Furui and
Faessler calculations and ours have approximately equally good fits to the
polarization, but
they predict a second ``zero-crossing" in the backward direction, which is not
seen in the
data.  Thus, both scalar and vector reaction mechanisms are needed for a good
fit to the data,
as we have argued above.
\medskip
\noindent
5.2  MESON-EXCHANGE MODELS
\medskip
Meson-exchange models have been proposed by Tabakin and Eisenstein [4],
Niskanen
[5], Kohno
and Weise [6], Timmermans et al. [7], LaFrance et al. [8] and Haidenbauer et
al.
[9].
These models differ in the types of K-mesons included in the exchange (K,
K$^*$,
K$^{**}$) and
in their initial and final state interactions.  The earlier calculations [4,5]
had a limited
amount of experimental data with which to compare their results.  Kohno and
Weise's
meson-exchange results were similar to their quark model calculations, which we
have discussed
above.  A comparison of our results with the later meson-exchange calculations
is given in
Fig.~5 for laboratory momenta of 1508 MeV/c and 1695 MeV/c.  Our fits are
better
than those of
LaFrance et al. and Haidenbauer et al., each of which fails to reproduce the
steep
forward rise in the differential cross section at 1695 MeV/c and predicts
oscillatory behavior
in the polarization at that momentum which is not seen in the experiment.  Our
fit to the
differential cross section and polarization is as good as that of the Nijmegen
group [7].
\bigskip
\centerline{\bf 6.  Depolarization}
\medskip
Haidenbauer et al. [1] have proposed that a measurement of the depolarization
parameter
$D_{nn}$ could be used to discriminate between quark and meson-exchange models.
This spin
observable measures the depolarization of the target in a direction $\hat n$
normal to the
reaction plane [21]
$$D_{nn} = {tr \left(\sigma^\Lambda_n \, M \, \sigma_n^p M^{\dag} \right) \over
tr \,
(MM^{\dag})} \eqno(7)$$
in which $M$ is the reaction matrix element.

Haidenbauer et al. noted that in Born approximation, the ``tensor-type"
interaction $\pmb\sigma_1
\cdot {\bf\hat r}\; \pmb\sigma_2 \cdot {\bf\hat r}$
which appears in meson exchange would
predict $D_{nn}
= -1$, whereas the vector quark interaction which includes the projection
operator on the
triplet state $P_1 = {1 \over 4} (3 + \pmb\sigma_1 \cdot \pmb\sigma_2)$
would predict
$D_{nn}
= 2/3$.  They found that a difference persisted even in the presence of initial
and final state interactions, as shown in Fig. 6, taken from reference [1]. The
meson-exchange calculations still predict $D_{nn} < 0$, and quark-model
calculations predict
$D_{nn} \gsim 0$.  We have extended this comparison to our SV quark model, and
find that
the different predictions persist.  In Fig. 7 we show calculations of $D_{nn}$
at 3 different
momenta.  We have considered 4 sets of parameters.  The first 3 are given in
Table 1: our
best global fit (both scalar and vector terms), our best global vector fit, and
our best
global scalar fit.  In addition we considered the best combined vector and
scalar fit at each
momentum.  In every case the depolarization calculated was $\gsim 0$, and
significantly
different from the meson-exchange model results of Haidenbauer et al.

To test the sensitivity of our calculation to the choice of initial state
interaction, we
repeated the calculation using the J\"ulich ``B" $\bar p p$ potential which
corresponds to the
solid curve of Fig. 6.  Our $\bar \Lambda \Lambda$ parameters were varied to
fit
the
experimental data at 1546 MeV/c.  Again, as shown in Fig. 8, the depolarization
remained $\gsim
0$.  The difference between quark-model and meson-exchange model predictions
for
$D_{nn}$
persists throughout the entire momentum range we have studied.
The measurement of $D_{nn}$ appears to be an excellent test of the models.

\centerline{\bf 7. Conclusion}
\smallskip
We have shown that an excellent fit to experimental data for the reaction $\bar
p p \rightarrow
\bar \Lambda \Lambda$ in the laboratory momentum region from 1436 to 1695 MeV/c
can be obtained
with a quark model that includes both scalar and vector terms in the reaction
mechanism.  This
model is sensitive to both initial and final state interactions. In order to
achieve good
agreement with experiment, a $\bar \Lambda \Lambda$ potential which differs
significantly from
that expected on the basis of SU(3) arguments is required.  The reaction takes
place
at distances for which quark effects are expected to be important, and our fits
at lower
momenta are comparable to the best of the meson-exchange calculations. At
higher
momenta our
quark model better fits the steep rise in $d\sigma/d\Omega$ at forward angles.
Further
comparison at higher momenta and for the production of $\bar \Sigma \Lambda,
\bar \Lambda
\Sigma$ and $\bar \Sigma \Sigma$ pairs will help to distinguish between these
models.  The
measurement of target depolarization $D_{nn}$ appears to be a strong test of
these
complementary pictures of strangeness production in $\bar p p \rightarrow
\bar \Lambda \Lambda$.
\medskip
\centerline{\bf Acknowledgments}
\smallskip
We wish to thank the members of the PS185
Collaboration for their interest in our calculations and the early
communication
of their
results. We have enjoyed helpful conversations with R. Eisenstein, D. Hertzog,
K. Holinde, K.
Kilian, V. Mull, F. Tabakin, and W. Oelert. This work is supported in part by
the U.S.
Department of Energy. One of us (M.A.A.) was also supported in part by the
National Science
Foundation under Grant No. PHY-9223618.
\medskip
\centerline{\bf References}
\smallskip
\item{1.}J. Haidenbauer, K. Holinde, V. Mull and J. Speth, Phys. Lett. (1992)
{\bf B 291},
223. \smallskip
\item{2.}P.D. Barnes et al., Phys. Lett. (1987) {\bf B189}, 249; Phys. Lett.
(1989) {\bf
B229},  432; Nucl. Phys. (1991) {\bf A526}, 575; H. Fischer, Ph. D. Thesis,
University of
Freiburg, Germany (1992); W. Oelert (private communication).
 \medskip
\item{3.}H. Dutz et al., CERN/SPSLC 92-53, SPSLC/1192 (1992).
\medskip
\item{4.}F. Tabakin and R.A. Eisenstein,
Phys. Rev. (1985) {\bf C31}, 1857.
\medskip
\item{5.}J.A. Niskanen, Helsinki preprint HU-TFT-85-28.
\medskip
\item{6.}M. Kohno and W. Weise, Phys. Lett. (1986) {\bf B179}, 15; Phys. Lett.
(1988) {\bf
B206},  584; Nucl. Phys. (1988) {\bf A479}, 433c.
\medskip
\item{7.}R.G.E. Timmermans, T.A. Rijken and J.J. deSwart, Nucl. Phys. (1988)
{\bf A479}, 383c;
Phys. Rev. (1992) {\bf D45}, 2288.
\medskip
\item{8.}P. LaFrance, B. Loiseau and R. Vinh Mau, Phys. Lett.  (1988) {\bf
B214}, 317; Nucl.
Phys.  (1991) {\bf A528}, 557.
\medskip
\item{9.}J. Haidenbauer et al.,
Phys. Rev. (1992) {\bf C45}, 931; J. Haidenbauer, K. Holinde, V. Mull and J.
Speth, Phys.
Rev. (1992) {\bf C46}, 2158.
\medskip
\item{10.}C.B. Dover and P.M. Fishbane, Nucl. Phys. (1984) {\bf B244}, 349.
\medskip
\item{11.}H. Genz and S. Tatur, Phys. Rev. (1984) {\bf D30}, 63; G. Brix, H.
Genz and S. Tatur,
Phys. Rev. (1989) {\bf D39}, 2054.
\medskip
\item{12.}P. Kroll, B. Quadder and W. Schweiger, Nucl. Phys. (1989) {\bf B316},
373.
\medskip
\item{13.}H.R. Rubinstein and H. Snellman, Phys. Lett. (1985) {\bf B165}, 187.
\medskip
\item{14.}S. Furui and A. Faessler, Nucl. Phys. (1987) {\bf A468}, 669.
\medskip
\item{15.}M. Burkardt and M. Dillig, Phys. Rev. (1988) {\bf C37}, 1362.
\medskip
\item{16.}M.A. Alberg, E.M. Henley and L. Wilets, Z. Phys. (1988) {\bf 331},
207; M.A. Alberg,
E.M. Henley, L. Wilets and P.D. Kunz, Nucl. Phys. (1990) {\bf A508}, 323c.
\medskip
\item{17.}M.A. Alberg, E.M. Henley, L. Wilets and P.D. Kunz, Nucl. Phys. (1993)
{\bf A560},
365. \medskip
\item{18.}M.A. Alberg, E.M. Henley and W. Weise, Phys. Lett. (1991) {\bf B255},
498.
\medskip
\item{19.}T. Ueda, Prog. Theor. Phys. (1979) {\bf 62}, 1670.
\medskip
\item{20.}M.M. Nagels, T.A. Rijken and J.J. de Swart, Phys. Rev. (1975) {\bf
D12}, 744; Phys.
Rev. (1977) {\bf D15}, 2547; Phys. Rev. (1979) {\bf D20}, 1633.
\medskip
\item{21.}J. Bystricky, F. Lehar and P. Winternitz, J. Phys. (Paris) (1978)
{\bf
39}, 1; P.
LaFrance and P. Winternitz, J. Phys. (Paris) (1980) {\bf 41}, 1391. \vfill
\eject
\centerline{\bf Figure Captions}
\medskip
\item{Fig.~1}Lowest order diagrams for $\bar p p \rightarrow \bar \Lambda
\Lambda$.
\medskip
\item{Fig.~2}Differential cross section and polarization at 1642 MeV/c for our
best global fit
parameters.  The experimental data is from PS185.
\medskip
\item{Fig.~3}A comparison of our best fits to the differential cross section
and
polarization
data at 1642 MeV/c using the vector (dashed line), scalar (dot-dashed line) or
combined
vector and scalar (solid line) reaction mechanisms.
\medskip \item{Fig.~4}Comparison of our calculations (solid line)
with other quark-based models:  the vector $``^3S_1"$ model of Kohno and Weise
[6] (dashed
line) and the scalar $``^3P_0"$ model of Furui and Faessler [14] (dot-dashed
line).
\medskip
\item{Fig.~5}Comparison of our global best fit calculation (solid line) with
the
meson-exchange
models of Timmermans et al.[7] (long-dashed line), LaFrance et al.[8]
(short-dashed line),
and Haidenbauer et al. [9] (dot-dashed line).
\medskip
\item{Fig.~6}Depolarization parameter $D_{nn}$ as calculated by Haidenbauer et
al. [1]. The
solid and dashed lines correspond to meson-exchange calculations with different
initial state
interactions.  The dot-dashed and dotted curves correspond to vector quark
model
calculations.
\medskip
\item{Fig.~7}Calculations of $D_{nn}$ for 3 different momenta using the global
fit parameters
of Table 1:  combined vector and scalar (solid line), vector (dashed line),
scalar (dot-dashed
line), and for the best combined vector and scalar fit parameters at each
momentum (dotted
line). \medskip
\item{Fig.~8}Calculation of $D_{nn}$ at 1546 MeV/c using the J\"ulich ``B"
$\bar
p p$ potential for our combined vector and scalar (solid line), vector (dashed
line) or scalar
(dot-dashed line) quark model mechanism.

\bye